\documentclass[12pt,preprint]{aastex}

\newcommand{\degree}{\hbox{$^\circ$}}

\shorttitle{Disintegration of Magnetic Flux in Decaying Sunspots}
\shortauthors{Kubo et al.}

\begin{document}

\title{Disintegration of Magnetic Flux in Decaying Sunspots as Observed
with the \textit{Hinode} SOT}

\author{M. Kubo\altaffilmark{1}, B. W. Lites\altaffilmark{1},
K. Ichimoto\altaffilmark{2}, T. Shimizu\altaffilmark{3},
Y. Suematsu\altaffilmark{2}, Y. Katsukawa\altaffilmark{2},
T. D. Tarbell\altaffilmark{4}, R. A. Shine\altaffilmark{4},
A. M. Title\altaffilmark{4}, S. Nagata\altaffilmark{5}}
\and
\author{S. Tsuneta\altaffilmark{2}}
\email{kubo@ucar.edu}

\altaffiltext{1}{High Altitude Observatory, National Center for
Atmospheric Research, P.O. Box 3000, Boulder, CO 80307. The National Center for
Atmospheric Research is sponsored by the National Science Foundation.} 
\altaffiltext{2}{National Astronomical Observatory of Japan, 2-21-1
Osawa, Mitaka, Tokyo 181-8588, Japan.} 
\altaffiltext{3}{Institute of Space and Astronautical Science, JAXA,
Sagamihara, Kanagawa 229-8510, Japan.} 
\altaffiltext{4}{Lockheed Martin Solar and Astrophysics Laboratory,
Building 252, 3251 Hanover Street, Palo Alto, CA 94304.} 
\altaffiltext{5}{Hida Observatory, Kyoto University, Takayama, Gifu
506-1314, Japan.}

\begin{abstract}
Continuous observations of sunspot penumbrae with the Solar Optical
 Telescope aboard \textit{Hinode} clearly show that the outer
 boundary of the penumbra fluctuates around its averaged position.
The penumbral outer boundary moves inward when granules appear in the
 outer penumbra.
We discover that such granules appear one after another while moving
 magnetic features (MMFs) are separating from the penumbral ``spines''
 (penumbral features that have stronger and more vertical fields than
 those of their surroundings). 
These granules that appear in the outer penumbra often merge with
 bright features inside the penumbra that move with the spines as they
elongate toward the moat region.
This suggests that convective motions around the penumbral outer
 boundary are related to the disintegration of magnetic flux in the
 sunspot.  
We also find that dark penumbral filaments frequently elongate into the moat
 region in the vicinity of MMFs that detach from penumbral spines.
Such elongating dark penumbral filaments correspond to nearly horizontal fields
 extending from the penumbra. 
Pairs of MMFs with positive and negative polarities are sometimes
 observed along the elongating dark penumbral filaments. 
This strongly supports the notion that such elongating dark penumbral
 filaments have magnetic fields with a ``sea serpent''-like structure. 
Evershed flows, which are associated with the penumbral
 horizontal fields, may be related to the detachment of the MMFs from the
 penumbral spines, as well as to the formation of the MMFs along the dark
 penumbral filaments that elongate into the moat region.
\end{abstract}


\section{INTRODUCTION}
Sunspot penumbrae consist of many dark and bright radial filaments.
The inclination of penumbral magnetic fields
fluctuates with azimuthal angle around the spot center
\citep{Degenhardt1991, Title1993, Lites1993,
Stanchfield1997, Westendorp2001a, Westendorp2001b, Bello2005, Langhans2005}. 
The bright penumbral filaments have magnetic fields that are more
vertical than those of the dark penumbral filaments in the outer
penumbra. 
The vertical and horizontal components of the penumbral fields have been
called a ``penumbral spine'' and a ``penumbral intra-spine,''
respectively \citep{Lites1993}.
Radial flows and radial motions are also observed in the penumbra.
The Evershed flow is a horizontal outward flow along the
horizontal component of the penumbral magnetic fields \citep{Degenhardt1991,
Title1993, Lites1993, Rimmele1995, Stanchfield1997, Westendorp2001a,
Westendorp2001b, Bellot2003, Bellot2004, Borrero2004, Borrero2005,
Cabrera2006, Cabrera2007, Cabrera2008}.
Evershed flows with supersonic velocities are sometimes inferred
around the outer boundary of the penumbra \citep{del2001}, and are
observed directly \citep{Shimizu2008b}.
Penumbral grains, which are patchy bright features, move toward the
umbra in the inner penumbra, and the penumbral grains in the
outer penumbra show both inward and outward motions
\citep[e.g.][]{Sobotka1999, Sobotka2001, Bovelet2003}.  
The Evershed flow appears to originate at the inward-moving
penumbral grains in the inner penumbra \citep{Rimmele2006, Ichimoto2007}.
Various models have been proposed to explain the complex magnetic and velocity
fields of penumbrae \citep[e.g.,][]{Solanki1993, Schlichenmaier1998,
Thomas2002, Spruit2006, Heinemann2007}.

Radial outward flows are dominant in the moat region that surrounds
mature sunspots. 
Moving magnetic features (MMFs) are small magnetic
elements that have sizes of typically
less than 2$\arcsec$ in the moat region 
\citep{Sheeley1969, Vrabec1971, Harvey1973}. 
MMFs mostly appear around the penumbral outer boundary and then move
almost radially outward in the moat region for mature and decaying sunspots
\citep{Harvey1973, Ryutova1998, Zhang2003, Sainz2005, Ravindra2006}. 
The lifetimes of MMFs range from a few minutes to 10 hr
\citep{Vrabec1974, Zhang2003, Hagenarr2005}.  
Those MMFs with longer lifetimes may reach the outer boundary of the
moat region. 
Recent observations suggest that MMFs located on lines extrapolated 
outward from the horizontal component of the penumbral magnetic fields into
the moat region correspond to the photospheric intersections of
undulating horizontal fields extending from the penumbra
\citep{Sainz2005, Cabrera2006, Cabrera2007, Kubo2007a}. 
This supports the ``sea serpent'' concept for MMFs, 
whereby the horizontal penumbral fields undulate into and out
of the solar surface
\citep{Harvey1973, Schlichenmaier2002}.
Many authors have proposed that the fluctuating penumbral fields and the
subsequent MMFs are produced by Evershed flows \citep[e.g.,][]{Ryutova1998,
Martinez2002, Thomas2002, Schlichenmaier2002, Zhang2003, Zhang2007b,
Cabrera2006}. 
On the other hand, MMFs located on the lines extrapolated from the
penumbral spines have magnetic fields that are more vertical than those
of their surroundings \citep{Kubo2007a}. 
Such MMFs appear to be formed as a result of the detachment of a
fraction of the magnetic field from the penumbral spines.

The net magnetic flux transported by all the MMFs exceeds
the flux-loss rate of the sunspot \citep{Martinez2002, Kubo2007a}. 
Hence, it has been suggested that the MMFs detached from
the penumbral spines alone are the agent that
removes magnetic flux from the sunspot
\citep[e.g.,][]{Shine2001, Martinez2002, Thomas2002, Weiss2004}.
Those MMFs might carry away an amount of magnetic flux 
sufficient to account for the flux loss of
the sunspot \citep{Kubo2007a}. 
If so, the formation of those MMFs that detach from the penumbral spines is
more important to the process of sunspot decay than that of other types
of MMFs. 

\citet{Kubo2007b} found that bright features (granules) appear where an
MMF is separating from the penumbral spine, but only one event was
examined in detail in that work.  
Using observations of two mature sunspots made by the Solar Optical
Telescope (SOT; Tsuneta et al. 2008; Suematsu et al. 2008; T. D. Tarbell et
al. 2008 in preparation; Ichimoto et al. 2008; Shimizu et al. 2008a)
aboard the \textit{Hinode} satellite \citep{Kosugi2007}, 
we build statistics of the comparison between
MMFs detaching from the spines and both bright and dark
features around the penumbral outer boundary.
The \textit{Hinode} SOT reveals the evolution of fine structures in both
the penumbra and the moat region from continuous observations with a
spatial resolution of $0.2\arcsec-0.3\arcsec$.

\section{OBSERVATIONS}
We focus on SOT observations of two mature sunspots that were located
near the disk center in order to be able to clearly identify the
penumbral spines and intra-spines with line-of-sight magnetograms.
These sunspots are typical decaying sunspots that have a nearly circular
shape. 
The data sets used in this study are summarized in
Table~\ref{data_summary}. 
We use the observations of the sunspot in NOAA Active Region (AR) 10933
to investigate temporal evolution in the G-band intensity and line-of-sight
magnetic fields in the photosphere over a large field of view. 
The temporal evolution of the vector magnetic field is obtained by the
\textit{Hinode} Spectropolarimeter (SP) for the sunspot in NOAA AR 10944. 
These data complement the G-band sequences with quantitative measures
of the vector field, but over a small field of view.

\subsection{Active Region NOAA 10933}
The filtergram (FG) of the SOT obtained sunspot images continuously in
NOAA AR 10933 with a 2 minute cadence from 16:14 to 24:00 on 2007 
January 4, and with a 7 minute cadence from 0:00 to 4:00 on 2007 January
5.  
The broadband filter imager (BFI) provided G-band 4305 {\AA} and Ca
{\small II} H 3968 {\AA} images over the full field of view (223$\arcsec
\times$112$\arcsec$). 
The narrowband filter imager (NFI) observed the intensity (Stokes
\textit{I}) and circular polarization (Stokes \textit{V}) at -120 {m\AA}
from the center of the photospheric line Fe {\small I} 6302.5 {\AA} with
the full field of view (328$\arcsec \times$164$\arcsec$). 
The filtergram images were binned 2$\times$2 pixels from the full
resolution in this observation.
The size of a binned pixel was 0.108$\arcsec$ for the BFI and
0.16$\arcsec$ for the NFI, respectively. 

The SP obtained a spatial distribution of the
full polarization state (Stokes \textit{I}, \textit{Q}, \textit{U}, and
\textit{V}) with the Fast Map mode from 18:40 to 19:43.
Having SP observations of the full polarization state permitted us to
infer the magnetic field vector and thermodynamic parameters in the photosphere.
In the Fast Map mode, the slit-scanning step was 0.297$\arcsec$, and a
pixel sampling along the slit was 0.320$\arcsec$.
The field of view was 297$\arcsec \times$ 164$\arcsec$.
The Stokes profiles of two magnetically sensitive Fe lines at 6301.5 and
6302.5 {\AA} were obtained simultaneously with a wavelength 
sampling of 21.6 {m\AA} and an integration time of 3.2 seconds for each
slit position. 
Repeated observations for an area of 3.9$\arcsec \times$
82$\arcsec$ were also carried out from 21:57 to 22:50. 
The evolution of the magnetic fields in this small area was reported in
\citet{Kubo2007b}.

\subsection{Active Region NOAA 10944}
The sunspot in NOAA AR 10944 was observed simultaneously 
with the FG and the SP for 6 hr from 11:51 on 2007
February 27. 
The BFI provided full-resolution images for the G-band and the Ca
{\small II} H line with a field of view of 56$\arcsec \times$ 56$\arcsec$. 
The NFI observed the Stokes \textit{I} and \textit{V} of the
photospheric line Fe {\small I} 6302.5 {\AA} in the shutterless mode, in
which  
continuous readout was performed for the central area of the CCD and the
outer parts of the CCD were masked. 
The polarization accuracy was better in the shutterless mode, but the
field of view was narrow (32$\arcsec \times$ 81$\arcsec$).
The time cadence of the BFI and NFI was 1 minute.

The SP repeatedly scanned the same region of 9.5$\arcsec \times$
 82$\arcsec$ with the Normal Map mode.
In the Normal Map mode, the spatial sampling was 0.149$\arcsec \times$
 0.160$\arcsec$, and the integration time for each slit position was 4.8
 seconds. 
The SP took about 5.5 minutes to obtain each map.

\section{DATA ANALYSIS}
\subsection{Filtergrams}
The standard calibrations, which were dark subtraction, flat-fielding, and bad
pixel correction, were performed for the G-band images with the BFI.
Only dark subtraction was applied for the Stokes \textit{I} images with
the NFI, because flat-fielding data for the NFI images were
not yet available for these data sets.  
Hereafter, we denote the Stokes \textit{V} image divided by the Stokes
\textit{I} image obtained at the same time as a ``line-of-sight
magnetogram.''  
The effect of flat-fielding on the line-of-sight magnetogram is small.  
Image cross-correlation allowed us to align
the calibrated G-band images to a reference image in order
to remove the drift due to the correlation tracker \citep{Shimizu2008a}
and the sunspot proper motion.
The image taken at the time of the midpoint of all the G-band images was
used as the reference. 
The line-of-sight magnetograms (Stokes \textit{V/I}) were aligned to the G-band
images that were taken at the time closest to the time of the NFI
observations via cross-correlation of the G-band to the Stokes
\textit{I} images.  

\begin{figure}
\epsscale{0.5}
\plotone{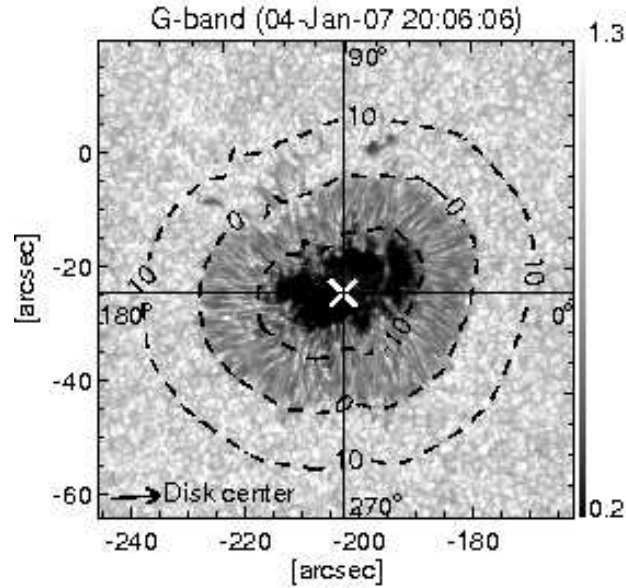}
\caption{Sunspot in NOAA AR 10933, observed with the \textit{Hinode} SOT.
The gray-scale bar at the right indicates intensity normalized to the 
mean intensity of the quiet area outside the sunspot. 
The cross corresponds to the geometrical center of the sunspot.
The solid lines show the position angle, measured counterclockwise from
 the solar west around the sunspot center.
The three dashed lines correspond to distances of -10$\arcsec$,
 0$\arcsec$, and 10$\arcsec$ from the penumbral outer boundary. 
The horizontal and vertical axes represent the positions with respect to
 the disk center.
} 
\label{20070104_sunspot}
\end{figure}

On the G-band image for NOAA AR 10933, Figure~\ref{20070104_sunspot}
shows the locations that we determined for the time-averaged penumbral
outer boundary, the geometrical center of the sunspot, and the reference 
frame for the position angles around the sunspot center.
The image averaged over all the G-band images was used to evaluate the
average intensities of the quiet area, the penumbral outer boundary, and
the central area of the sunspot. 
The quiet-area intensity ($I_0$) was determined by a Gaussian fit 
to the distribution of the intensities for pixels located well outside
the sunspot.
We spatially smoothed the averaged G-band image with a box of 5$\arcsec
\times$ 5$\arcsec$ to remove local fluctuations due to penumbral fine
structure. 
Let $I_G$ be the smoothed G-band intensity.  
We then define the average penumbral outer boundary as the locus of $I_G
= 0.87 I_0$.  
In NOAA AR 10933, the sunspot center was chosen as the center of gravity
(COG) of the $I_0 - I_G$ image for the entire sunspot. 
For NOAA AR 10944, we use the COG of the $I_0 - I_G$ image for the umbra
($I_G < 0.4$) only because the images did not cover the entire penumbral
area. 

We used only the southern part for the sunspot in NOAA AR 10933, because
a largest bubble was stably located at the top of the NFI images
(T. D. Tarbell et al. 2008, in preparation).
The southern part of the sunspot appeared to be clean.

\subsection{Spectropolarimeter}
We performed the following calibrations for the Stokes profiles observed
with the SP: 
(1) dark subtraction and flat-fielding, (2) the polarization calibration
induced by the optical elements in the SOT, (3) the correction in
the direction of spectral dispersion, (4) merging of two orthogonal
polarizations simultaneously measured with the left and right segments of
the CCD, (5) the correction for spectral curvature, (6) compensation for
residual crosstalk \textit{I} $\rightarrow$ \textit{QUV}, (7) the
correction for the orbital shift in the wavelength and slit directions
due to the thermal flexure of the Focal Plane Package
(FPP; T. D. Tarbell et al. 2008, in preparation), and (8) an intensity
correction due to the slit width variation along the slit.  
After the calibration, vector magnetic fields were derived with a
least-squares fitting (``Stokes inversion'') that assumed a 
Milne-Eddington representation of the atmospheric stratification
(T. Yokoyama et al. 2008, in preparation). 
Of the 13 parameters obtained with the Stokes inversion,
we consider only the field strength ($|{\bf{B}}|$), the inclination angle 
relative to the line of sight ($\gamma$), and the filling factor ($f$).
The magnetic field maps thus obtained were aligned to the G-band images
taken at the time closest to the midpoint of the SP maps, using
cross-correlation of the SP continuum intensity map with the 
BFI G-band images.

One must resolve the 180$\degree$ ambiguity of the azimuth angle in
order to calculate the inclination with respect to the normal to
the solar surface.  
We first select the line-of-sight azimuth closest to the potential field
computed using the line-of-sight component of the magnetic field as a
boundary condition.
Then we interactively determined the azimuth to reduce discontinuities
of azimuth and inclination angles by using the AZAM utility
\citep[written in IDL by P. Seagraves;][]{Lites1995},
which allowed us to display and select the two alternatives for the
azimuth angle.
Both the sunspots were located near the disk center, so
any errors in the inclination arising from the wrong choice of 
the ambiguous azimuth will amount to less than 15$\degree$.  
In any event, an erroneous choice of azimuth is rare for symmetric
sunspots such as those studied here.

\section{RESULTS}
\subsection{Magnetic Fields of Bright/Dark Features in the Outer Penumbra}
\begin{figure}
\epsscale{0.85}
\plotone{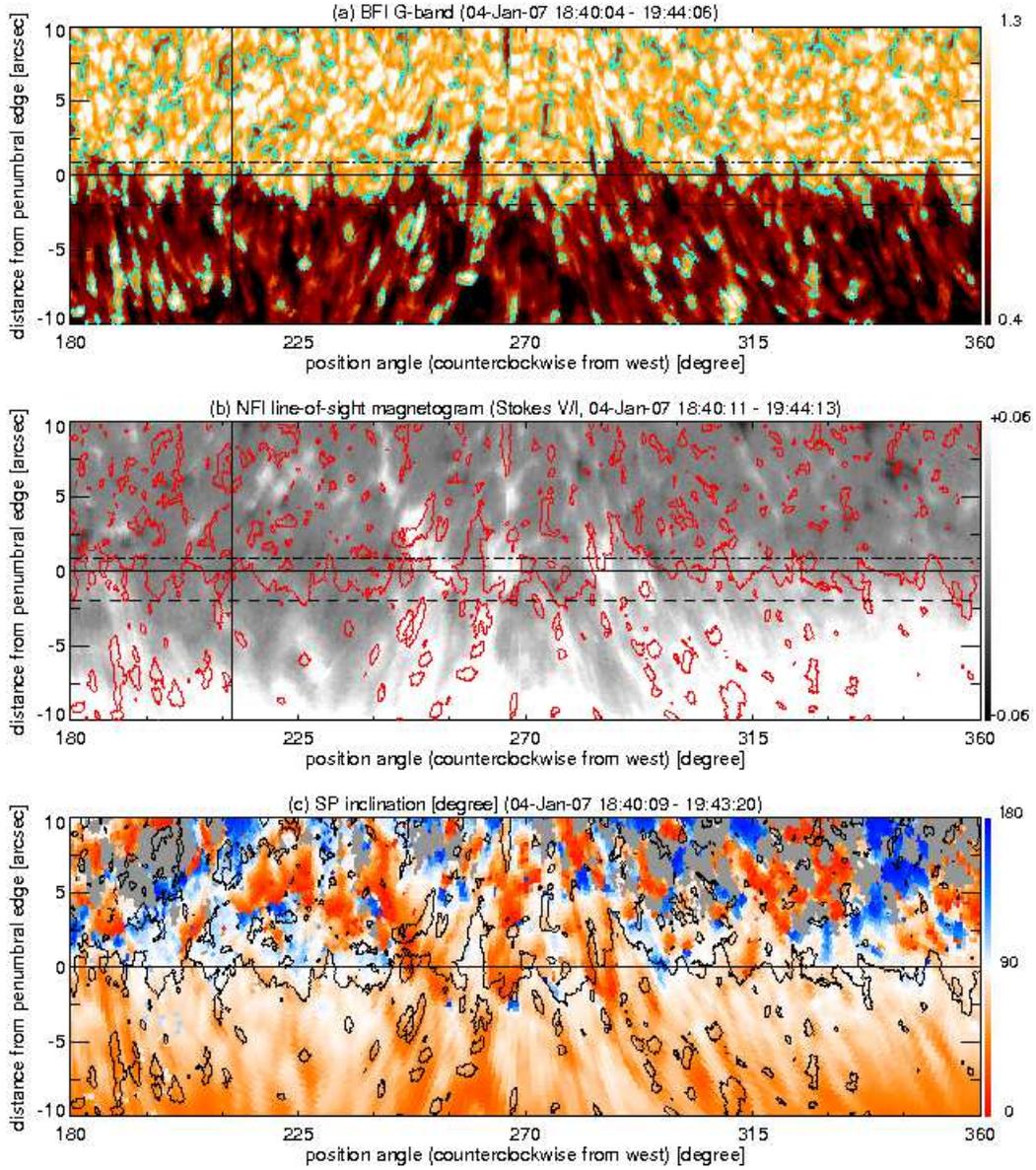}
\caption{(\textit{a}) G-band image, (\textit{b}) line-of-sight magnetogram,
 and (\textit{c}) magnetic field inclination.
The vertical axis shows the radial distance from the penumbral outer
 boundary, and the horizontal axis shows the position angle around the
 sunspot center.  
These panels are made from the observations of NOAA AR 10933 on 2007
 January 4.
Each pixel in panels \textit{a} and \textit{b} was aligned to the
 time and position closest to those of each pixel in panel \textit{c}. 
The G-band intensity is normalized to the quiet-area intensity. 
The contours represent a G-band intensity of 0.87.
White (black) indicates positive (negative) polarity in panel \textit{b}.
An inclination of 90$\degree$ corresponds to magnetic fields directed
 perpendicular to the local vertical, and inclinations of 0$\degree$ and
 180$\degree$ represent magnetic fields directed away from the solar
 surface and magnetic fields directed toward the surface, respectively. 
The horizontal solid line shows the averaged outer boundary of the
 penumbra.
The horizontal dashed and dash-dotted lines in panels \textit{a} and
 \textit{b} represent distances of 2$\arcsec$ inside and 1$\arcsec$
 outside the outer penumbral boundary, respectively. 
The vertical solid line in panels \textit{a} and \textit{b} shows the
 position angle of 212$\degree$. 
} 
\label{20070104_lFOV}
\end{figure}

The penumbra consists of many radial structures, and MMFs usually move
radially outward from the outer penumbral edge.
Figure~\ref{20070104_lFOV} displays on the
vertical axis the
radial distance from the averaged penumbral outer boundary, and on the
horizontal axis, the position angle around the sunspot center.  
The contours in Figure~\ref{20070104_lFOV} enclose features brighter in
the G band than $0.87I_0$. 
Many of the features enclosed by the contours in
Figure~\ref{20070104_lFOV}\textit{a} may be found in the middle and
outer penumbra. 
Some of these correspond to stronger positive line-of-sight
fields (Fig.~\ref{20070104_lFOV}\textit{b}).  
Furthermore, Figure~\ref{20070104_lFOV}\textit{c} demonstrates that the
enclosed radial structures have magnetic fields that are more vertical
than those of their surroundings.
The line-of-sight direction is nearly normal to the solar surface in
this case. 
Therefore, we can assign the white and gray areas in the outer penumbra
of Figure~\ref{20070104_lFOV}\textit{b}
as the vertical (spine) component and the horizontal (intra-spine)
component of the penumbral magnetic fields, respectively.
We can also identify the patchy magnetic elements in
Figure~\ref{20070104_lFOV}\textit{b} that have intense white 
and black areas in the moat region as MMFs with vertical fields. 

Bright features in the G-band images usually have a radially elongated
structure. 
\citet{Langhans2005} showed, using the finest spatial resolution
($0.2\arcsec$) ever achieved, that bright penumbral filaments have
more vertical magnetic fields than those of dark penumbral filaments in
the outer penumbra.
We confirm this relationship using spectropolarimetric measurements 
with a spatial resolution of $0.3\arcsec$.

\subsection{MMFs Separating from the Penumbra}
\begin{figure}
\epsscale{1.0}
\plotone{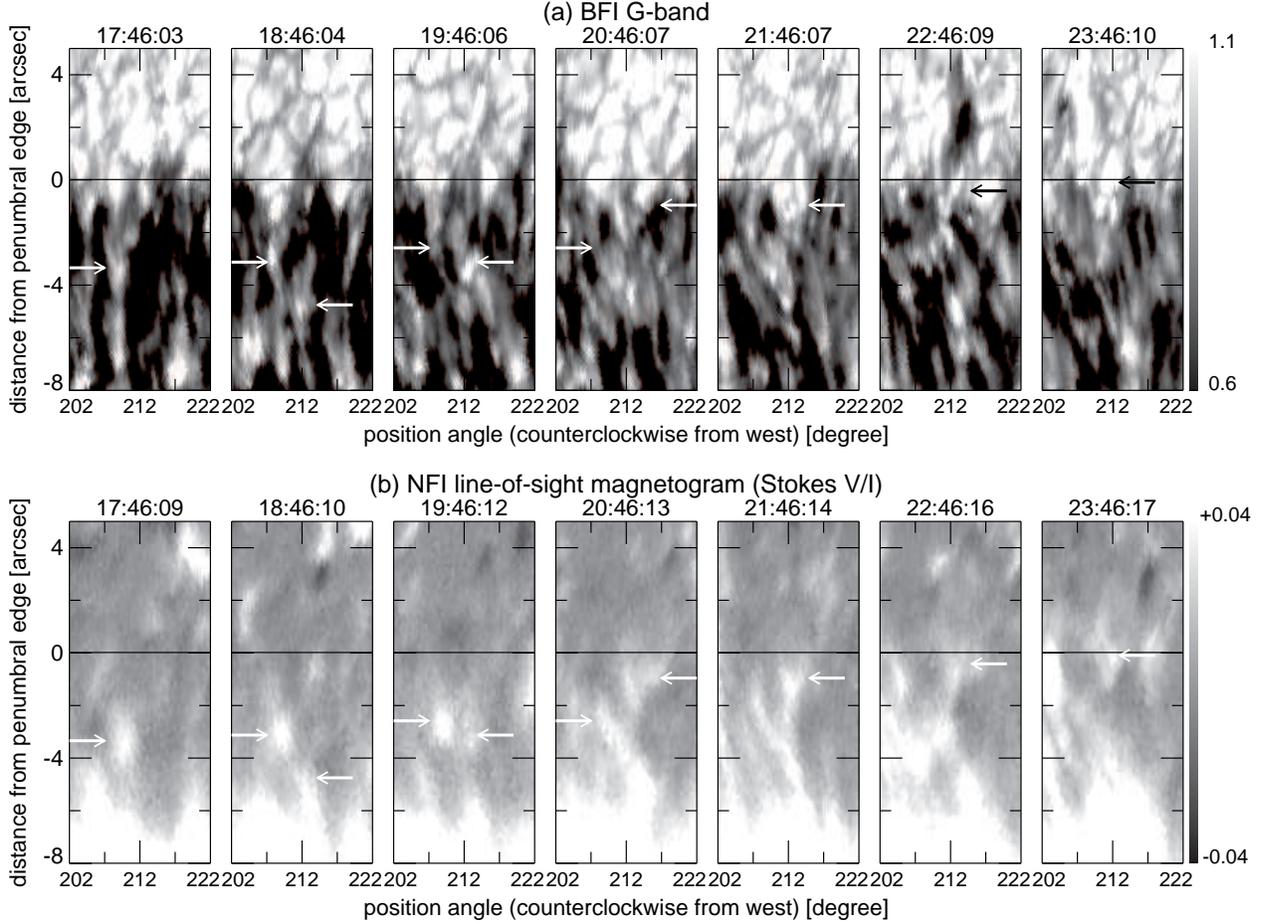}
\caption{Time series of (\textit{a}) G-band images and (\textit{b})
 line-of-sight magnetograms for position angles in the range
 202$\degree-222\degree$ on 2007 January 4.
The horizontal axis represents the position angle around the sunspot
 center.
The vertical axis represents the radial distance from the averaged
 penumbral outer boundary.
The horizontal solid line shows the averaged outer boundary of the penumbra.
The G-band intensity is normalized to the quiet-area intensity. 
White indicates positive polarity and black indicates negative polarity
 in panel \textit{b}. 
See the text for the interpretations of the arrows in the panels.
}
\label{20070104_sFOV}
\end{figure}

We find a clear correspondence between MMFs separating from the
penumbral spines and convective motions in the outer penumbra.
Figure~\ref{20070104_sFOV} shows a time series of G-band images and
line-of-sight magnetograms around the position angle of 212$\degree$
(indicated by the vertical lines in Fig.~\ref{20070104_lFOV}).
A penumbral spine appears in the frame at 18:46, as shown by the arrow
pointing to the left, and it elongates radially outward. 
In the same period, a bright elongated structure appears in the G-band
image, and its brightest area also moves outward with the elongating
penumbral spine. 
The elongating spine connects with an existing magnetic element
(indicated by the arrow pointing to the right) from 19:46 to 20:46, and
the outer part of the elongating spine becomes large in this period. 
The outer part of the elongating spine is gradually separating from
the main body of the spine, and then this part becomes an MMF in the
frame at 23:46.
We find that granules appear in the outer penumbra when the outer edge
of the elongating spine reaches the penumbra outer boundary (see the
21:46 frame). 
Such granules continuously appear one after another while the MMF is
detaching from the spine. 

\begin{figure}
\epsscale{1.0}
%\plotone{f4.eps}
\plotone{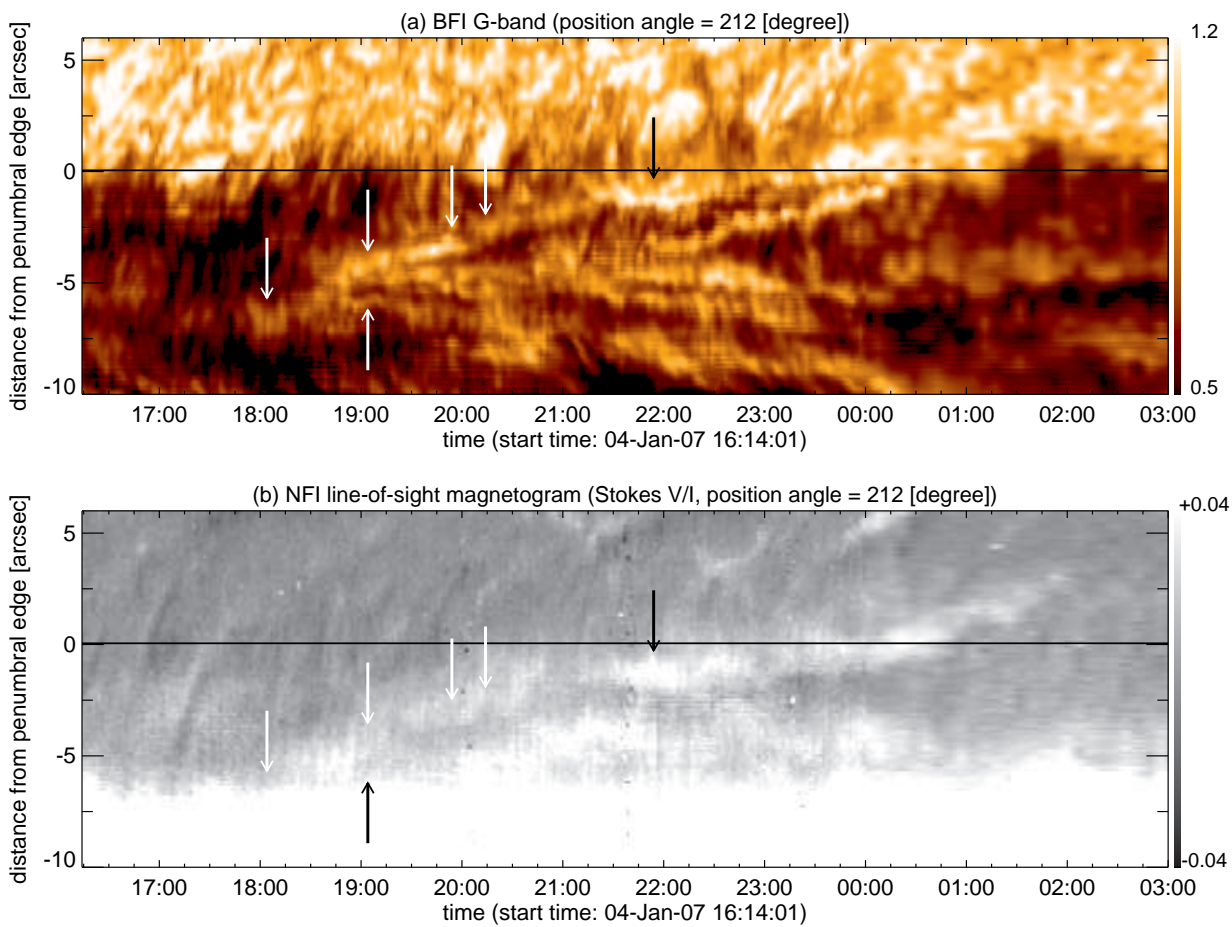}
\caption{Space vs. time plots along the line at the position angle of
 212$\degree$ for (\textit{a}) the G-band intensity and
 (\textit{b}) the line-of-sight magnetic field.
The position angle of 212$\degree$ is shown by the vertical line 
in Fig.~\ref{20070104_lFOV}.
These figures present results of two consecutive observations, differing
 mainly by their cadence. 
The cadence is 2 minutes from 16:14 to 24:00 on 2007 January 4 and is 7
 minutes from 0:00 to 3:00 on 2007 January 5. 
The G-band intensity is normalized to the quiet-area intensity. 
White indicates positive polarity and black indicates negative polarity
 in panel \textit{b}. 
See the text for the interpretation of the arrows in the panels.
} 
\label{20070104_st_radial}
\end{figure}

Figure~\ref{20070104_st_radial} shows the temporal change of the G-band
intensity and the line-of-sight magnetic field along a radial line at the
position angle of 212$\degree$. 
Two kinds of magnetic elements with outward motion can be seen from the
outer penumbra to the inner moat region, as reported in
\citet{Zhang2007a}: fuzzy, small magnetic elements move quickly, and 
large magnetic elements move slowly.
There is no significant G-band bright feature in the outer and middle
penumbra, and only the fuzzy magnetic elements are visible there in the
first 2 hr. 
Then, a bright feature appears around 18:00 on January 4 and moves
toward the moat region with an elongating large magnetic element, which
corresponds to the elongating penumbral spine shown in
Figure~\ref{20070104_sFOV}.  
The bright feature divides into two features around 19:00: one moves
toward the umbra (the upward arrow in
Fig.~\ref{20070104_st_radial}\textit{a}), and the other moves outward
with the elongating spine to become an MMF (downward arrow). 
The bright feature moving outward disappears around 20:00, and another
bright feature subsequently appears 20 minutes later at a position similar 
to where the previous feature disappeared.
This new bright feature also moves outward with the elongating spine.
The outer edge of the spine arrives at the penumbral outer boundary
around 22:00, and 
as a result of the appearance of granules in the outer penumbra, 
the penumbral outer boundary retracts toward the umbra relative
to its average position.
The bright feature moving with the elongating spine merges with the
appearing granules.  
The outer part of the elongating spine is completely separated from the
penumbra around 00:00, and this part still moves outward as an MMF.
The elongating speed of the spine averaged over 6 hr after its
appearance is about 0.2 km s$^{-1}$, which is similar to the horizontal
velocities of MMFs that have vertical magnetic fields with the same
polarity as that of the sunspot around the penumbral outer boundary
\citep{Kubo2007a}.  
There are many bright features moving inward and outward in the penumbra
before 00:00 on January 5, but most of the bright features and the elongating
spine disappear from the outer penumbra after the MMF separates from
the penumbral spine. 

\begin{figure}
\epsscale{1.0}
\plotone{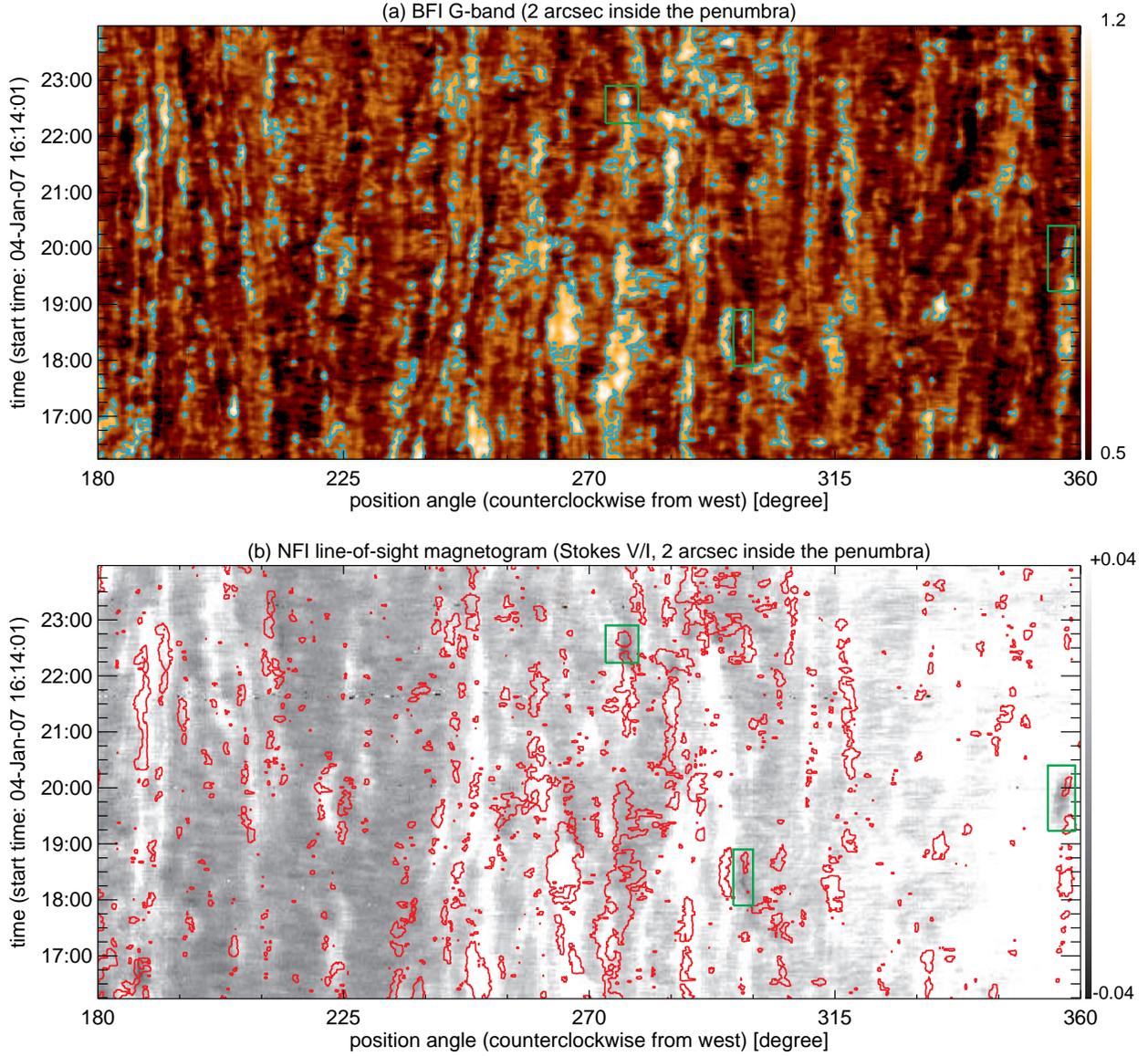}
\caption{Space vs. time plots along the dashed line at 2$\arcsec$ inside the
 penumbral outer boundary in Fig.~\ref{20070104_lFOV} for (\textit{a})
 the G-band intensity and (\textit{b}) the line-of-sight magnetic field. 
The contours represent a G-band intensity of 0.87 with respect to the
 quiet-area intensity.  
White indicates positive polarity and black indicates negative polarity
 in panel \textit{b}. 
The green boxes show bright features located at the weak black areas in
 panel \textit{b}.}
\label{20070104_st_inside}
\end{figure}

The relationship between the MMF separating from the elongated spine and
convection in the outer penumbra is demonstrated above for only the
radial line at the position angle of 212$\degree$.
To confirm that this relationship is common all around the
penumbral outer boundary, we compare the temporal change in 
the G-band intensity to the line-of-sight magnetic field at 2$\arcsec$
inside the penumbral outer boundary for position angles ranging from 
180$\degree$ to 360$\degree$ in Figure~\ref{20070104_st_inside}. 
There are many bright streaks in
Figure~\ref{20070104_st_inside}\textit{a}. 
These bright streaks represent the granules appearing in the outer penumbra or
the penumbral bright features moving outward.
On the other hand, the positive magnetic features in
Figure~\ref{20070104_st_inside}\textit{b} mostly correspond to the
elongating spines or the MMFs detaching from the spines.
We confirm that the G-band bright streaks are generally located at the
elongating spines or at the detaching MMFs, as shown by the contours in
Figure~\ref{20070104_st_inside}.
This tendency is more clearly seen in the limb-side penumbra (the left-hand
side of Fig.~\ref{20070104_st_inside}) than in the center-side
penumbra due to the projection effect. 

Some bright features are not located at positive magnetic features but
at negative magnetic features with weak magnetic fields (darker
features than their surroundings; see the green boxes in
Fig.~\ref{20070104_st_inside}).  
The bright features in the outer penumbra are related to formation
of the MMFs that have the polarity opposite to that of the sunspot.

\subsection{Dark Penumbral Filament Elongating to the Moat Region}
\begin{figure}
\epsscale{1.0}
\plotone{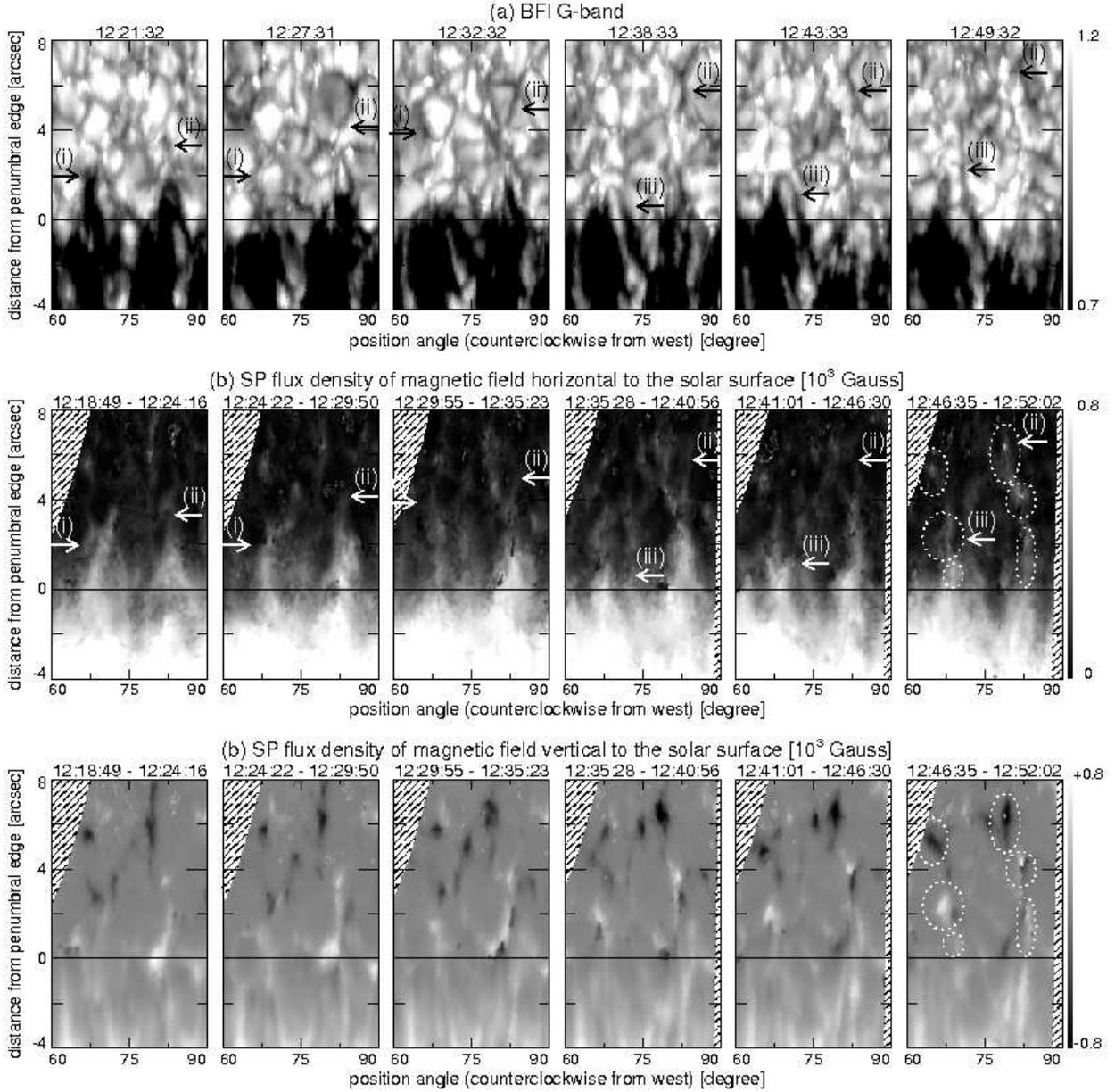}
\caption{Time series of (\textit{a}) the G-band images, (\textit{b})
 the magnetic flux density of the horizontal magnetic fields, and
 (\textit{c}) the magnetic flux density of the vertical magnetic fields.
The horizontal axis represents the position angle around the sunspot
 center.
The vertical axis represents the radial distance from the penumbral
 outer boundary.
These panels were obtained from the observations of NOAA AR 10944 on 2007
 February 27.
The magnetic flux densities of the horizontal field
 [$f|{\bf{B}}|\sin(\gamma)$] and the vertical field
 [$f|{\bf{B}}|\cos(\gamma)$] are derived from the SP observation, where
 the field strength is $|{\bf{B}}|$, the inclination angle with respect
 to the local vertical is $\gamma$, and the filling factor is $f$. 
The hatched areas with oblique lines represent the area that is out of the SP
 field of view.
See the text for the interpretation of the arrows and circles in the
 panels. 
}
\label{20070227_sFOV}
\end{figure}

The outer boundary of the penumbra fluctuates around its average
position, and dark penumbral filaments often elongate into the moat
region.
Such dark filaments correspond to the areas with strong horizontal
fields that extend from the penumbra.
The arrows in Figure~\ref{20070227_sFOV} indicate the evolution of 
three dark, elongating
penumbral filaments with strong horizontal fields in NOAA AR 10944.  
The longer dark penumbral filaments extend to the outer moat region.
The MMFs with the polarity opposite to that of the sunspot are located
at the outer edge of the filamentary structure with strong horizontal fields.
This is consistent with the result of \citet{Kubo2007b}.
The dotted circles in the rightmost frames of
Figures~\ref{20070227_sFOV}\textit{b} and \ref{20070227_sFOV}\textit{c} 
indicate that two MMFs with positive polarity and
two MMFs with negative polarity are aligned along each of two filamentary
structures that have a strong horizontal field (the middle circles each
contain two MMFs that have opposite polarities). 
This is clear evidence that bipolar MMFs correspond to the
intersections of the solar surface with serpentine horizontal fields
extending from the penumbra.
In this case, each serpentine horizontal field line consists of both
$\Omega$-loops and U-loops (see Fig.~\ref{summary_fig}\textit{b}).  
This means that whether bipolar MMFs form $\Omega$-loops or U-loops
depends on which segments of the extended penumbral fields are chosen,
as well as on the number of intersections. 
The bipolar MMFs with a U-loop may correspond to a localized dip of
the magnetic canopy, as proposed by \citet{Zhang2003}.

\begin{figure}
\epsscale{1.0}
\plotone{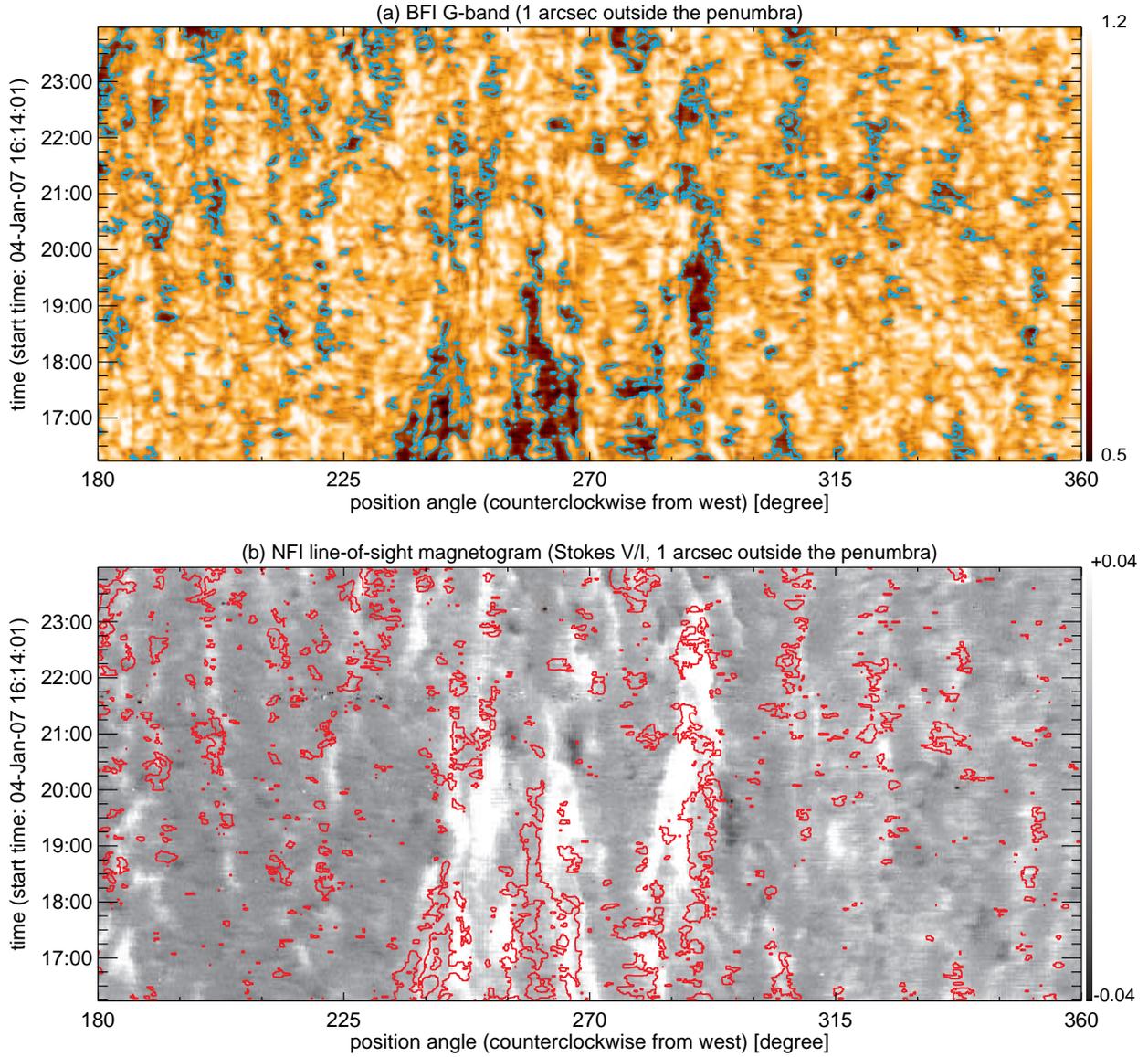}
\caption{Same as Fig.~\ref{20070104_st_inside}, but for space vs. time
 plots at 1$\arcsec$ outside the penumbral outer boundary (the
 dash-dotted line in Fig.~\ref{20070104_lFOV}).} 
\label{20070104_st_outside}
\end{figure}

Figure~\ref{20070104_st_outside} shows the temporal change in the G-band
intensity and the line-of-sight magnetic field at 1$\arcsec$ outside the
penumbral outer boundary of NOAA AR 10933.   
The dark penumbral filaments elongating to the moat region are indicated
by the dark features in Figure~\ref{20070104_st_outside}\textit{a}.
The appearance and disappearance of the long positive magnetic features
in Figure~\ref{20070104_st_outside}\textit{b} are mainly due to the
outward motion of MMFs that detach from the penumbral spines.
The elongating dark filaments (red contours in
Fig.~\ref{20070104_st_outside}\textit{b}) are usually located at the
areas that have nearly horizontal magnetic fields with respect to the
solar surface (gray or weak white areas in
Fig.~\ref{20070104_st_outside}\textit{b}). 
We find that most of the elongating dark filaments are adjacent to the
MMFs that have the same polarity as the sunspot. 
This means that the dark penumbral filaments elongate near the MMFs that
have the same polarity as the sunspot when such MMFs separate from the
penumbra. 
One example of such a dark penumbral filament and an MMF can be seen at
22:46 in Figure~\ref{20070104_sFOV}.

\section{SUMMARY AND DISCUSSION}
\begin{figure}
\epsscale{0.8}
\plotone{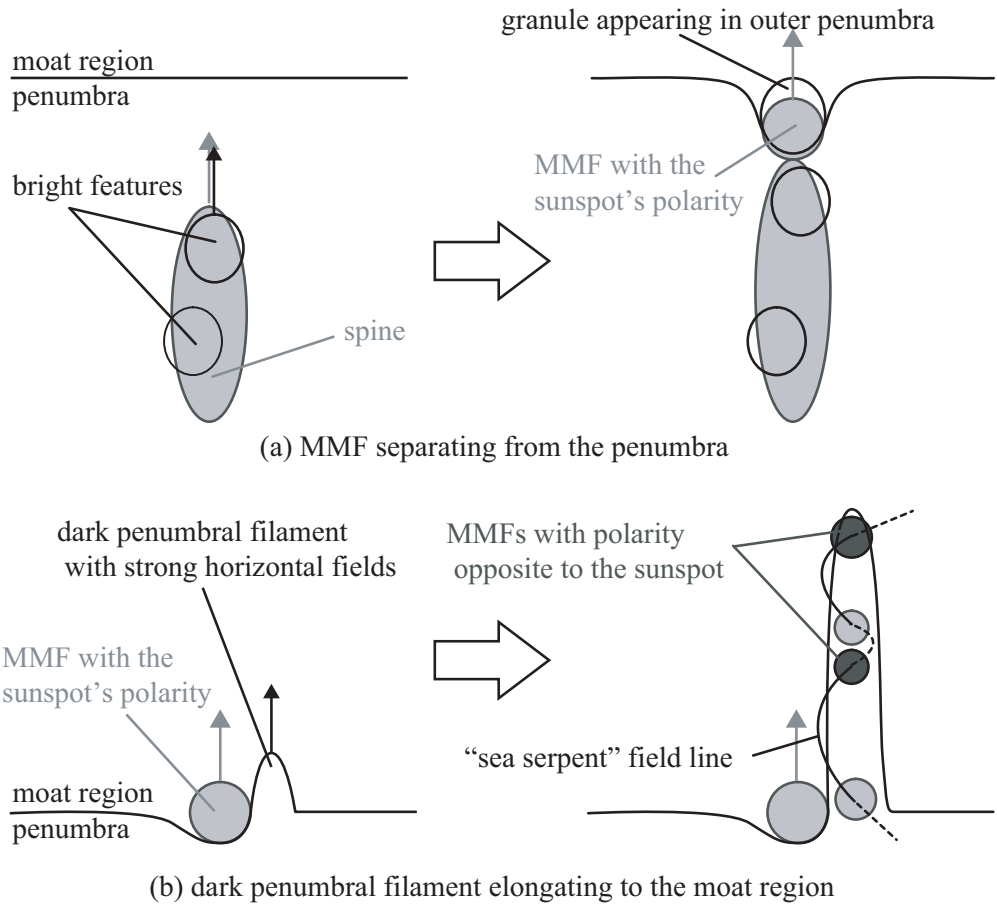}
\caption{Schematic illustration for a summary of the observations around the
 penumbral outer boundary.}
\label{summary_fig}
\end{figure}

The SOT allows us to perform continuous observations with high spatial
resolution for a duration longer than the typical lifetime of MMFs.
The results regarding flux removal from the sunspot in this study are
summarized as follows (see also Fig.~\ref{summary_fig}):

\begin{enumerate}
\item G-band bright features in the outer penumbra are located at 
      the elongating spines or at the MMFs detaching from the spines.
      G-band bright features moving both outward and inward can
      be found in the
      middle and outer penumbra.
      The penumbral spines elongate toward the moat region in concert with the
      outward motions of the penumbral bright features.
\item Granules appear in the outer penumbra at locations where MMFs that
      have the same polarity as the sunspot separate from the penumbral spines.
      Such granules often merge with bright features moving along
      the elongating spines.
\item Dark penumbral filaments frequently elongate into the moat
      region, and longer ones reach the outer moat region. 
      The elongating dark penumbral filaments coincide with
      filamentary structures that have strong horizontal fields that
      extend outward from the penumbra. 
      We present clear examples of bipolar MMFs indicating a serpentine
      field with multiple intersections with the solar photosphere.
      MMFs with positive and negative polarities are
      located along the filamentary structures with strong horizontal
      fields. 
\item The MMFs that are detached from the penumbral spines are usually
      observed in the neighborhood of the elongating dark penumbral filaments. 
\end{enumerate}

The total flux transport rate by the MMFs that are detached from the
penumbral spines is greater than the flux-loss rate of the sunspot
\citep{Kubo2007a}.
Therefore, items 1 and 2 above confirm that convection in the outer
penumbra is related to the flux removal of the sunspot \citep{Kubo2007b}.
However, the magnetic energy density of the penumbral fields is larger
than the kinetic energy density of the penumbral bright features at the
photospheric surface.
Converging and downward flows are inferred in the subsurface convection
zone around a stable sunspot \citep{Zhao2001}, and such flows might be
essential to the formation and sustenance of sunspots \citep{Meyer1974,
Parker1979, Parker1992}. 
The appearance of the bright features moving outward in the outer
penumbra suggests subsurface upwelling and diverging flows.
Such flows would work against the stabilization of the sunspot, and
recent MHD simulations showed that magnetic flux is carried away from the
spot by subsurface diverging flows \citep{Heinemann2007}. 
\citet{Meyer1974} suggested that small flux tubes diffuse out
into the moat region by small scale convection beneath the sunspot.
The fact that the bright features and granules appear in the outer
penumbra at the location of the elongating spines (that subsequently
become MMFs) supports this idea.

Another possibility is that Evershed flows detach penumbral flux
and advect it into the moat region. 
Evershed flows with supersonic velocities are often observed
around the penumbral outer boundary \citep{Shimizu2008b}. 
Supersonic flows are also
observed in connection with horizontal fields extending from the penumbra
\citep{Kubo2007b}. 
The kinetic energy density of such supersonic flows is larger than the magnetic
energy density of the magnetic fields in the outer penumbra.
Nevertheless, the \textit{Hinode} SP observations of vector magnetic
fields with 0.3$\arcsec$ resolution have confirmed that the Evershed
flows are radial outward flows along those penumbral filaments that have
nearly horizontal fields \citep{Ichimoto2007}. 
Detachment of the MMFs is observed when the penumbral horizontal fields
elongate into the moat region.  
This means that Evershed flows are related not only to the formation of
the bipolar MMFs located along the extending penumbral horizontal
fields, but also to the detachment of the MMFs from the penumbral
spines.   
The Evershed flows originate at penumbral bright features that are
moving inward in the inner penumbra \citep{Rimmele2006, Ichimoto2007},
and the detachment of MMFs from the penumbra is observed with penumbral
bright features that are moving outward in the outer penumbra.
This suggests the possibility that both Evershed flows and the detachment
of MMFs originate from the activity of convection below the sunspot
penumbra.  
The relationship between Evershed flows and sunspot decay also should be
investigated using Doppler and magnetic field measurements of both high
spatial and high temporal resolution.

\acknowledgments
The authors would like to acknowledge the late Professor T. Kosugi and
all the members of the \textit{Hinode} team for their work to realize a
successful mission.
We also thank T. Yokoyama and M. Shimojo for their help in deriving
the magnetic field vector.
R. Centeno is also thanked for the co-alignment method between 
the G-band images and the SP maps.
\textit{Hinode} is a Japanese mission developed and launched by
ISAS/JAXA, with NAOJ as domestic partner and NASA and STFC (UK) as
international partners. 
It is operated by these agencies in cooperation with ESA and
NSC (Norway). 
This work was partly carried out at the NAOJ \textit{Hinode} Science
Center, which is supported by the Grant-in-Aid for Creative Scientific
Research ``The Basic Study of Space Weather Prediction'' from MEXT,
Japan (Head Investigator: K. Shibata), generous donations from Sun
Microsystems, and NAOJ internal funding.

\clearpage

\begin{deluxetable}{cccccccc}
\tabletypesize{\scriptsize}
\tablecolumns{8}
\tablewidth{0pc}
\tablecaption{Data Sets Used in This Study\label{data_summary}}
\tablehead{
\colhead{} &\colhead{} &\colhead{} &\colhead{}
 &\colhead{} &\colhead{Field of view} &\colhead{Cadence} &\colhead{}\\
\colhead{NOAA AR} &\colhead{Location} &\colhead{Date} &\colhead{Period}
 &\colhead{SOT Observation} &\colhead{(arcsec)} &\colhead{(minutes)} &\colhead{Figure}
}
\startdata
10933 &S02, E12 &2007 Jan 04&16:14-24:00 &BFI\tablenotemark{a} (G-band) &223$\arcsec \times$
 112$\arcsec$ &2 &\ref{20070104_sunspot}, \ref{20070104_lFOV},
 \ref{20070104_sFOV}, \ref{20070104_st_radial}, 
 \ref{20070104_st_inside}, \ref{20070104_st_outside} \\ 
      &         &            &16:14-24:00 &NFI\tablenotemark{b} (Stokes \textit{I}, \textit{V})
 &328$\arcsec \times$ 164$\arcsec$ &2 &\ref{20070104_lFOV},
 \ref{20070104_sFOV}, \ref{20070104_st_radial}, 
 \ref{20070104_st_inside}, \ref{20070104_st_outside} \\ 
      &         &            &18:40-19:43 &SP\tablenotemark{c} &297$\arcsec \times$ 164$\arcsec$
 &63\tablenotemark{d} &\ref{20070104_lFOV} \\
      &         &2007 Jan 05&00:00-03:58 &BFI (G-band) &223$\arcsec \times$
 112$\arcsec$ &7 &\ref{20070104_st_radial} \\ 
      &         &            &00:00-03:58 &NFI (Stokes \textit{I}, \textit{V})
 &328$\arcsec \times$ 164$\arcsec$ &7 &\ref{20070104_st_radial} \\ 
10944 &S01, E13 &2007 Feb 27 &11:57-17:54 &BFI (G-band) &56$\arcsec \times$
 56$\arcsec$ &1 &\ref{20070227_sFOV} \\ 
      &         &            &11:51-17:51 &SP &9.5$\arcsec \times$
 82$\arcsec$ &5.5\tablenotemark{d} &\ref{20070227_sFOV} \\
\enddata
\tablenotetext{a}{Broadband filter imager} 
\tablenotetext{b}{Narrowband filter imager} 
\tablenotetext{c}{Spectropolarimeter} 
\tablenotetext{d}{Cadence for one map} 
\end{deluxetable}

\end{document}